\documentclass[journal=nalefd,manuscript=letter,layout=onecolumn,email=false]{achemso}

\usepackage[version=3]{mhchem} 
\usepackage{mathpazo}

\author{Chin Wang}
\affiliation[Beijing Institute of Technology]
  {School of Physics, Beijing Institute of Technology, Beijing, China 100081}
\alsoaffiliation[Hunan University of Arts and Science]
  {Department of Physics, Hunan University of Arts and Science, Changde, China 415000}
\author{Luogen Deng}
\affiliation[Beijing Institute of Technology]
  {School of Physics, Beijing Institute of Technology, Beijing, China 100081}

\title[]
  {Electrically Controlled Plasmonic Lasing Resonances with Silver Nanoparticles Embedded in Amplifying Nematic Liquid Crystals}

\abbreviations{IR,NMR,UV}

\begin{document}
\begin{spacing}{1.0}

\begin{abstract}
We demonstrate an electrically controlled coherent random lasing with silver nano-particles dispersed in a dye-doped nematic liquid crystal (NLC), in which external electric field dependent emission intensity and frequency-splitting are recorded.
A modified rate equation model is proposed to interpret the observed coherent lasing, which is a manifestation of double enhancements, caused by the plasmon-polariton near-fields of Ag particles, on the population inversion of laser dye molecules and the optical energy density of lasing modes. The noticeable quenching of lasing resonances in a weak applied field is due to the dynamic light scattering by irregular director fluctuations of the NLC host, which wash out the coherent interference among different particle palsmon-polariton fields. This provides a proof to support that the present lasing resonances are very sensitive to the dielectric perturbations in the host medium and thus are likely associated with some coupled plasmonic oscillations of metal nanoparticles.
\end{abstract}


\subsection{Introduction}
Since random lasing was demonstrated in a series of disordered gain materials\cite{lawandy1994laser,cao1999random}, the interest in this kind of novel mirrorless lasers has grown very rapidly because of their extensive application prospect in the fields of display, target identification, biomedicine and photonic devices. The necessary condition for achieving a typical random lasing is that the disordered laser material is required to support coherent or incoherent feedback and meantime provide a sufficient optical gain for the seed photons to be lasing. In those of original experimental investigations it was some dielectric particles with high positive permittivities that were chose to fabricate random lasers\cite{lawandy1994laser,cao1999random,garcia2007photonic,Noginov2005lasers}, where the multiple scattering mechanism is essential for establishing optical feedback. Until recently one of inspiring breakthroughs was made by X. Meng and co-workers when they observed coherent random lasing from dye-doped polymer films embedded with very tiny Ag nanoparticles\cite{meng2008random,meng2009coherent}. In these extremely transparent disordered samples, the scattering mean free paths of light are far larger than the wavelengths as well as the sample thicknesses, which rules out any evident multiple scattering mechanism of confining light. Such efforts suggest that the localized plasmon-polariton fields of Ag nanopartiles exhibit very competitive capabilities to improve and tailor the stimulated emission characteristics of the surrounding gain media. Moreover the particle plasmon-polariton can be easily controlled by engineering the particle size and shape, changing the dielectric and chemical environments, and modulating the electromagnetic coupling with other adjacent particle partners. However, to fully exploit the merits of metal-nanoparticle-based random lasers in designing optical devices is hindered sometimes by the lack of effective dynamic control over the emission. In view of the rich optical properties of nematic liquid crystals on birefringence and light scattering due to the orientations and fluctuations of their directors\cite{gennes1993physics}, in particular, which can be controlled by external field, environment temperature and anchoring force at interfaces, we hope to incorporate some tunable NLC environments into a plasmon-type random laser and tailor the optical properties of such device. This desire is also driven by another unknown question that whether the intrinsic scattering properties of a NLC itself can be directly utilized to offer an effective optical-feedback for random lasing or, at least, provide an enhancement effect for plasmon-type random lasers.

In this letter, we report an experimental and theoretical  investigation for the random lasing behavior of a dye-doped NLC embedded with dilute silver nanoparticles under optical pumping excitation. We found that above associated thresholds profound lasing resonances emerge in this random laser system if the director orientational fluctuation of the NLC was suppressed by an applied electric field, however, no random lasing signal was probed in the presence of a sufficient director fluctuation, i.e., introducing the dynamic light scattering due to the NLC director fluctuation yet destructs the coherent feedback already provided by the included Ag nanoparticles and then extinguishes the random lasing. By modulating the director fluctuation via a tunable external electric field, we have achieved a switching of this Ag-nanoparticle-based random lasing. Furthermore, we as well recorded an intriguing frequency-splitting ($\sim$2.5nm) of the lasing resonances and its slight shrink as increasing the strength of the external electric field. Based on a rate equation theory of laser, we point out that the plasmon-polarition near-fields of the Ag nanopaticles contribute to an overlapped local-field enhancement on both the population inversion of laser dye (at pumping frequency) and the energy density of laser field (at lasing frequencies), and thus account for the coherent lasing spikes on the emission spectra. As we expected, the measured dependence of random laser intensity on the external electric field can be qualitatively interpreted according to the physics of the dynamic light scattering of NLCs.

\subsection{Experimental details}
The semi-closed sample cell that consists of a pair of ITO-conductive glass separated by a 40 $\mu m$ spacer and glued at three sides was in advance prepared. The disordered material to create random laser is an ultrasonically-dispersed mixture that consists of a commercial NLC ($n_{o}=1.52$ and $n_{e}=1.72$ at 610 nm and 20 $^{0}C$), dodecanethiol-passivated Ag nanoparticles with nearly monodisperse dimension, and 0.5 mMol/L laser dye DCM. The Ag nanoparticles were synthesized by wet chemical synthesis\cite{shankar2010wet} and characterized with a transmission electron microscope (JEM-1011, JEOL) operating at 100 kV accelerating voltage. Figure 1a shows the TEM image of the Ag nanoparticles whose average particle size is about 10 nm in diameter. The number density of the Ag nanoparticles was arranged to be $\sim 8.85\times 10^{17} /m^3$. This disordered mixture was infiltrated into the prepared sample cell to form the random laser sample, which we hereafter refer to as NLC-Ag-DCM sample. As shown in Figure 1b, such a sample was optically pumped by the second harmonics of a mode-locked neodymium-doped yttrium aluminum garnet (YAG) laser ($\lambda =532$ nm, 10 Hz repetition rate, 10 ns pulse duration) in order to perform random laser experiments. The pump beam polarized along x-axis direction was focused on the sample through a cylindrical lens to form an excitation stripe with the length of 1.8 mm and a variable width from 0.20 to 0.28 mm. The emission along the stripe was measured by a fiber optic spectrometer (AvaSpec-2048) with its spectral resolution of ~0.2 nm.

\begin{figure}[h]
\begin{center}
\includegraphics[width=5.0cm]{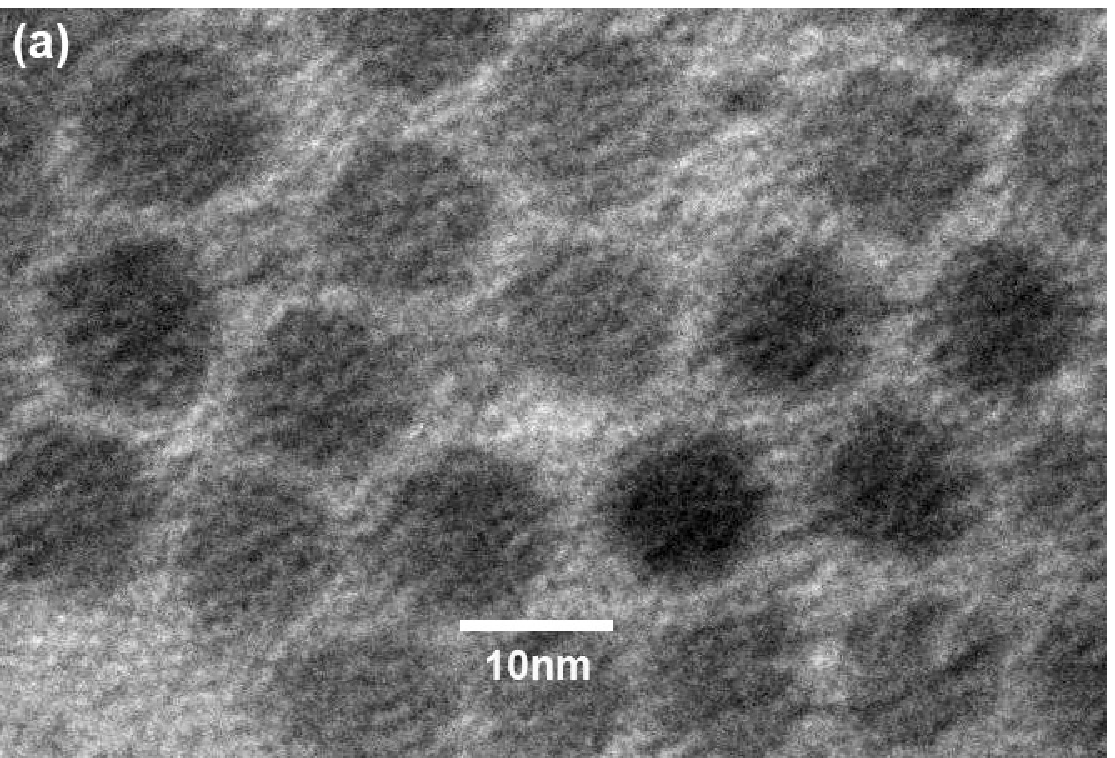}
\,
\includegraphics[width=6.4cm]{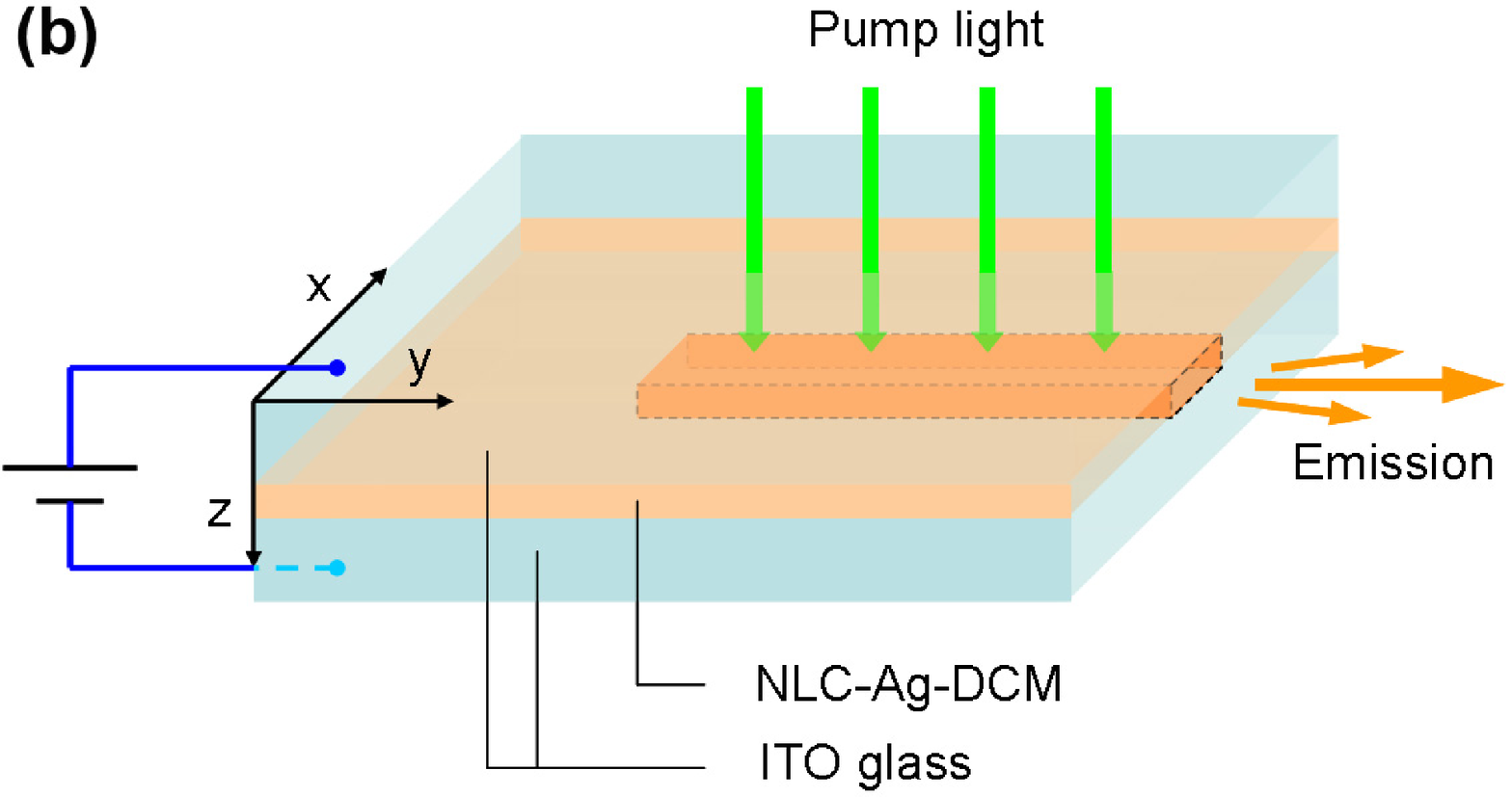}
\,
\includegraphics[width=4.0cm]{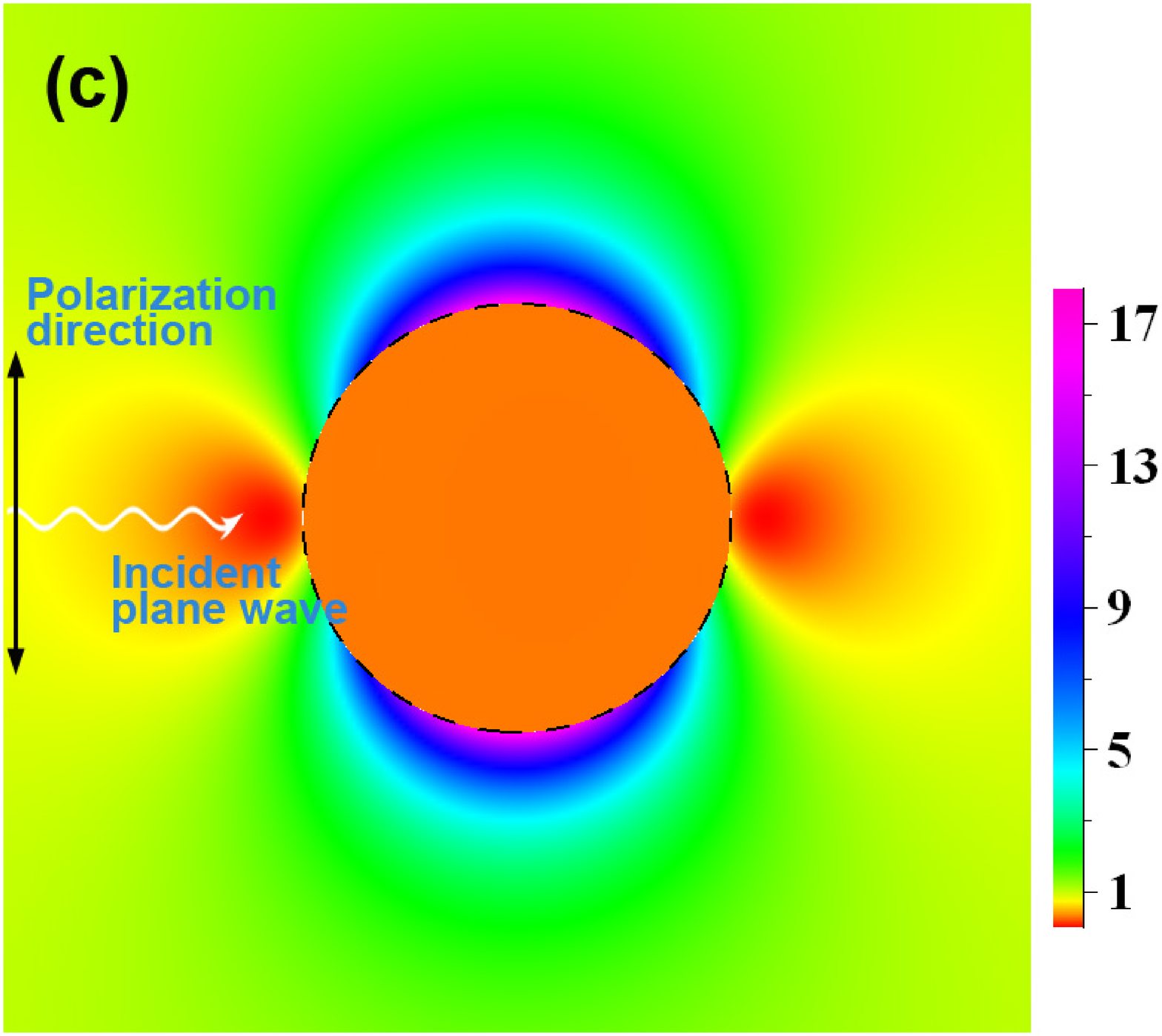}
\end{center}
\caption{(a) The TEM image shows the morphology and size of silver nanoparticles. (b) Schematic illustration of the configuration of NLC-Ag-DCM sample and the arrangement of excitation-detection.  (c) Energy density distribution of a particle plasmon-polariton mode excited by incident photons at 610nm. The dashed black circle labels the particle surface.}
\label{}
\end{figure}

\subsection{Result discussions and analyses}

\begin{figure}[h]
\begin{center}
\includegraphics[width=8.0cm]{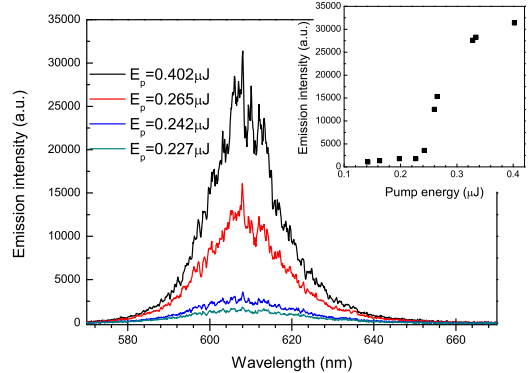}
\end{center}

\caption{Main panel indicates the evolution of emission spectra as a function of pump energy per pulse,
and the insert gives the dependence of emission intensity on the pump energy.
The emission spectra were obtained from a pumped strip with $\sim$1.8mm length and $\sim$0.28mm width, under fixed electric field of 0.75 V/$\mu m$.}
\label{pump dependence}
\end{figure}

To study the emission behavior, we initially exerted an external electrical field of 0.75 V/$\mu m$ across the sample and adjusted the pump energy. Beside that the emission spectra exhibited broad spontaneous emission features at weak pumping, a narrow peak appeared at 608 nm once the pumping energy exceeded a critical threshold value, as seen in \ref{pump dependence}, of which the insert gives the intensity of this peak as a function of pumping energy, yielding an input-output curve with a pronounced threshold characteristic of lasers. However the small ratio of the laser peak to the enhanced background of spontaneous emission suggests that there is still a minority of excited DCM molecules contributing to the stimulated emission in the laser mode. As increasing the pump energy, more distinguishable spectral lines with full widths of $\sim 0.2$ nm at half maximum (FWHM) emerged on the backgrounds of enhanced spontaneous emission spectra. Although it is a clear signal of coherent lasing resonances, the physics behind these laser spikes is by no means an obvious mechanism. As we will see in the following analysis, the observed coherent lasing resonances at least are associated with two enhancement effects caused by the particle plasmon-polariton near-fields of Ag nanoparticles: one is the enhanced excitation rate of dye molecules at pumping frequency, and the other is the enhanced stimulated transition probability of them at lasing frequencies.

In all of possible optical feedback mechanisms we first evaluate the contribution of the multiple scattering of light by the randomly distributed Ag nanoparticles. It is expected that, when the fluctuation of NLC's director is suppressed by a strong applied electrical field, the NLC-Ag-DCM sample becomes almost transparent and its residual scattering property is dominated by the buried silver nanoparticles. The extinction cross section of a 10 nm diameter silver nanosphere embedded in the NLC host (average refractive index $\sim$1.59) is calculated by Mie theory to be $\sigma_{e} \sim 6.91 \times 10^{-19} m^{2}$ for $\sim$610 nm emission light. Together with the information of the particle density, $\rho \sim 8.85 \times 10^{17} /m^{3}$, the transport mean free path $l_{t}$ of emission light in the NLC-Ag-NPs sample can be estimated via the formula $l_{t} = 1/(\rho \sigma_e) \sim 1.64$ m, which is much larger than the wavelength and the sample size, indicating that this NLC-Ag-DCM sample actually operates in ballistic regime as it was exposed in a considerable external electric field. This consideration on transport mean free path also excludes a Anderson localization mechanism\cite{john1984electromagnetic,Anderson1985} for the optical feedback to establish random lasing.

Since there is no apparent multiple-scattering-like optical feedback to account for the observed random lasing from this NLC-Ag-DCM sample, we turn to the possible influences resulting from the plasmonic oscillations of the Ag nanoparticles. An UV-visible extinction spectrum for an extremely dilute suspension of such Ag nanoparticles (embedded in toluene solvent which has a refractive index very close to that of the NLC host at the random lasing frequencies) was obtained from a spectrophotometric measurement, to compare with the absorption and fluorescence spectra of DCM molecules (in the absence of Ag nanoparticles), as well as one typical emission spectrum of the random lasing, as all shown in Figure 3. The extinction spectrum of Ag-NPs exhibits a surface plasmon resonance band centred at 425nm corresponding to the collective oscillations of the 5s electrons of silver atoms. Like in all conventional dye lasers, the observed random lasing spikes emerge at wavelengths where the dye absorption is practically absent, but the fluorescence emission is still strong. Furthermore, this lasing behavior actually survived in a region far beyond the plasmon resonance peak of individual Ag nanoparticles and thus avoided a notable absorption loss caused by them. However, because both the natural fluorescence emission of dye molecules and the plasmon-excitation of individual metal particles exhibit none of discrete frequency-dependences, the essential coherent feedback mechanism responsible for the sharp lasing resonances is still hidden from us.  A potential interpretation is that the coherent optical feedback in our NLC-Ag-DCM sample is probably caused by some high-Q and coupled plasmonic oscillations of Ag nanoparticles' collections, instead of being simply attributed to the plasmon resonances of individual Ag nanoparticles. The speculation is that, in a collection of metal nanoparticles with small separations, the opposite polarization charges at the surfaces of two adjacent particles attract each other across the gap between them, thus weakening the Coulomb restoring force inside each nanoparticle, and this gives rise to a large red-shift of the coupled plasmon resonance with respect to the single-particle plasmon resonance\cite{lamprecht2000metal,quinten2001color,su2003interparticle,nordlander2004plasmon,fromm2004gap,romero2006plasmons,biring2008light}.
To validate this hypothesis, we introduce a dielectric perturbation in the NLC host, by removing the pre-applied electric field to release the director fluctuation without changing the configuration of particle distribution, and found that the lasing resonances indeed disappear as the disordered dynamic fluctuation of the NLC director washes out coherent interference among different particle plasmon-polariton fields.
\begin{figure}[h]
\begin{center}
\includegraphics[width=8.0cm]{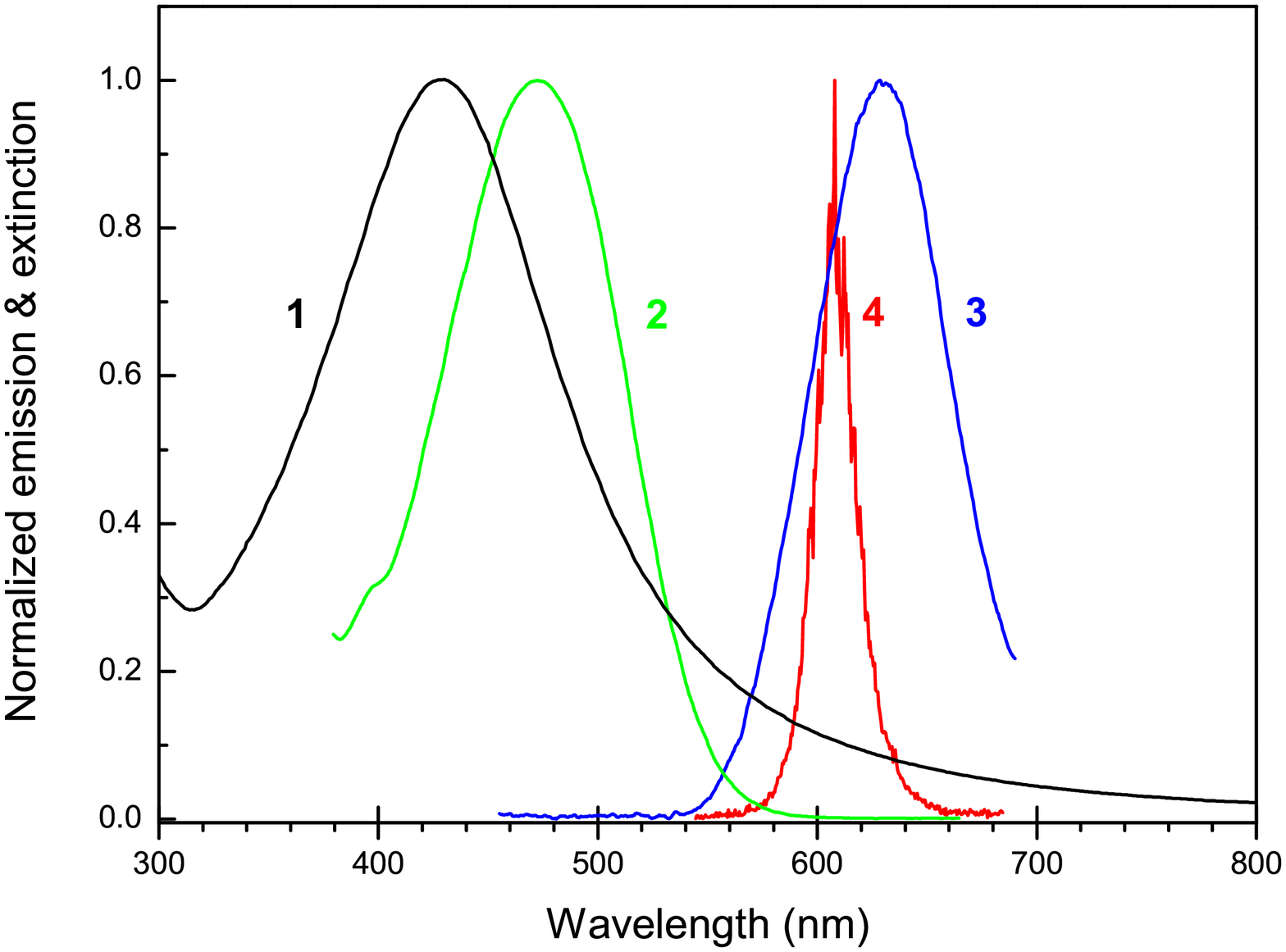}
\end{center}
\caption{Extinction spectrum of Ag nanoparticles of 10 nm diameter (1), absorption (2) and emission (3) spectra of DCM,
and one typical emission spectrum of the NLC-Ag-DCM sample (4).}
\label{Spectroscopic analysis}
\end{figure}

A plasmonic oscillation in fact is competent to provide optical feedback for lasing. Once a plasmon-polariton of nano-sized metal particles is excited by external irradiation, it allows compact storage and feedback of the injected energy into the localized surface fields near the regions where net charges and/or oscillating currents accumulated. These localized surface fields thus exhibit extreme enhancements in their strengths compared to that of the initial irradiation field, and is bound to dramatically improve the stimulated absorption and emission efficiencies of dye molecules if placed there. The improved stimulated absorption efficiency at pumping frequency contributes an enhanced pumping rate, while the improved stimulated emission efficiency at laser frequencies leads to a dominant weight of the coherent photons in total emission relative to those incoherent ones resulting from spontaneous emission.

To establish an intuitive understanding of the localized-field enhancement effect due to metal particle plasmons, we investigate the electric field around one Ag nanoparticle embedded in its NLC host, assuming an identical permeability, $\mu \approx \mu_0$, of the silver particle and the surrounding NLC host.  A time-harmonic light wave with frequency $\omega$ propagating in the NLC host is described by the associated Maxwell wave equation
\begin{equation}\label{}
\nabla  \times \left( {\nabla  \times {\bf{E}}} \right) - \omega ^2 \mu \varepsilon _ \bot  {\bf{E}}  = \omega ^2 \mu \varepsilon _\Delta  {\bf{n}}\left( {{\bf{n}} \cdot {\bf{E}}} \right).
\end{equation}
Here, $\varepsilon_{\Delta}=\varepsilon_{\|}-\varepsilon_{\bot}$ denotes the difference between the two principal dielectric constants $\varepsilon_{\parallel}$, for the polarizations parallel to the optical axis of the NLC medium, and $\varepsilon_{\perp}$, perpendicular to this axis; the unit vector \textbf{n} dictates the \emph{average} director orientation of NLC and coincides with the ${\bf{e}}_z $ direction in the discussed experimental configuration (Figure 1b). By contrast, inside the Ag nanoparticle the matched time-harmonic light wave just obeys the vector wave equation
\begin{equation}\label{}
\nabla  \times \left( {\nabla  \times {\bf{E}}} \right) - \omega ^2 \mu \varepsilon _{{\rm{Ag}}} {\bf{E}} = 0
\end{equation}
with an isotropic and metallic dielectric function $\varepsilon_{\rm{Ag}}$.

For a simple model we assume that the Ag nanoparticle has a spherical shape characterized by its radius R, and the total laser field about the particle is driven by some incoming photons emitted from the DCM molecules doped in the NLC host. Specifically, we now arrange an appropriate local spherical-polar-coordinate system ($r, \theta, \phi$), symmetrically fixed on the Ag nanosphere so that these driving photons perform as a linearly polarized plane wave incident along the polar axis, with the polarization direction dictated by radial unit vector ${\bf{e}}_p  \equiv {{\bf{e}}_r } |_{\varphi  = 0,\;\theta  = \pi /2}$. The scattered field together with the incident driving field determine the details of the enhanced near-field of this Ag nanosphere. In the following discussion we introduce a wave number parameter $k = \omega \sqrt {\varepsilon _ \bot  \mu }$. By treating the
$ \varepsilon _\Delta   \ll \varepsilon _ \bot $ as a small perturbation parameter, the zero-order approximation yet the primary part of the light field outside the metal sphere can be obtained directly according to Mie scattering theory, which shows that the total outer field can be well-approximated by the driving field overlapping an infinite series of spherical vector harmonics\cite{bohren1998absorption}, i.e.,
\begin{equation}\label{plasmonfield}
{\bf{E}} = E_0 e^{ikr\cos \theta } {\bf{e}}_p  + \sum\limits_{n = 1}^\infty  {E_n \left( {ia_n {\bf{N}}_{e1n}^{(H)}  - b_n {\bf{M}}_{o1n}^{(H)} } \right)}  + o(\varepsilon _\Delta  ),
\end{equation}
in which $E_n  = i^n E_0 (2n + 1)/(n^2  + n)$, and $E_{0}$ is the amplitude of the incident driving field; the vector spherical harmonicses \textbf{M} and \textbf{N} are defined, in terms of spherical Hankel functions and standard spherical harmonics, as
\begin{subequations}
\begin{eqnarray}
{\bf{M}}_{o1n}^{(H)}  &=& \nabla  \times \left( {{\bf{r}}h_n^{(1)} (kr)P_n^1 (\cos \theta )\sin \varphi } \right), \\
{\bf{N}}_{e1n}^{(H)}  &=& k^{ - 1} \nabla  \times \left[ {\nabla  \times \left( {{\bf{r}}h_n^{(1)} (kr)P_n^1 (\cos \theta )\cos \varphi } \right)} \right],
\end{eqnarray}
\end{subequations}
which at far-field region have some asymptotic forms of out-going spherical waves, compatible with the boundary condition of lasing modes; and the standard Mie scattering coefficients $a_n$ and $b_n$ are given by
\begin{subequations}
\begin{eqnarray}
\label{MeiCoefficient}
a_n &=& \frac{{m^2 j_n (m\rho )[\rho j_n (\rho )]' - j_n (\rho )[m\rho j_n (m\rho )]'}}{{m^2 j_n (m\rho )[\rho h_n^{(1)} (\rho )]' - h_n^{(1)} (\rho )[m\rho j_n (m\rho )]'}} ~ ,\\
b_n &=& \frac{{j_n (m\rho )[\rho j_n (\rho )]' - j_n (\rho )[m\rho j_n (m\rho )]'}}{{j_n (m\rho )[\rho h_n^{(1)} (\rho )]' - h_n^{(1)} (\rho )[m\rho j_n (m\rho )]'}} ~ ,
\end{eqnarray}
\end{subequations}
where $\rho  = kR$ is a dimensionless particle size, $m = ( \varepsilon _{{\rm{Ag}}} / \varepsilon _ \bot )^{1/2}$ is a relative refractive index,
and the superscript '$\prime$' denotes a derivative of the function with respect to the argument in parentheses.

After the necessary parameters are determined as follow:  $ \varepsilon _ \bot   = \varepsilon _0 \, n_{\rm{o}}^2  = 2.31 \,\varepsilon _0 $,
$ \varepsilon _{{\rm{Ag}}}  = \varepsilon _0 \, n_{{\rm{Ag}}}^2 $, and  $ n_{{\rm{Ag}}}  = 0.06 + 4.152\,i $ extracted from the experimental data\cite{johnson1972optical} for $\sim610nm$ wavelength, a straightforward calculation of the scattering coefficients from formulas (5) shows that, for the investigated 10nm-diameter Ag nanosphere, in \ref{plasmonfield} only the terms characterizing the incident driving filed and the electric dipole scattering field $ {\bf{N}}_{e11}^{(H)} $ are relatively significant, and thus we have
\begin{equation}\label{near field}
{\bf{E}} \approx E_0 \left( {e^{ikr\cos \theta } {\bf{e}}_p  - {\textstyle{3 \over 2}}a_1 {\bf{N}}_{e11}^{(H)} } \right),
\end{equation}
from which the enhancement of energy density (time average value) for near-field can be well-approximated by
\begin{equation}\label{}
\frac{{{\bf{E}}^*  \cdot {\bf{E}}}}{{E_0^2 }} \approx 1 + \frac{{3{\mathop{\rm Im}\nolimits} \{ a_1 \} }}{{k^3 }}\frac{{(1 - 3\sin ^2 \theta \cos ^2 \varphi )}}{{r^3 }} + \frac{{9|a_1 |^2 }}{{4k^6 }}\frac{{(1 + 3\sin ^2 \theta \cos ^2 \varphi )}}{{r^6 }}.
\end{equation}
As shown in figure 1c, this plasmon-polariton field at a lasing resonance exhibits an up to 17-times enhancement of the energy density on the charge-accumulated surface. Although such a plasmon-polariton is excited by photons assumed in a plane wave form, more general excitations should be expanded with plane waves bases, and the averaging over all propagation directions and polarizations does not change the basic physical picture of some enhanced near-fields. Because this localized-field enhancement mechanism is also suitable for the pumping light, a similar analysis reveals that a linearly polarized pump beam at 532nm can excite a surface electric field with the energy density enhanced by more than 25 times, which results in an enhanced pumping rate of dye molecules.

We assume that the dye molecules in the NLC-Ag-DCM sample can be modeled by an open two-level system, with the upper and lower levels being allowed to exchange their populations with a thermal reservoir, independently\cite{boyd2003nonlinear}. In addition to the stimulated emission and absorption channels induced by coherent laser field, the decay of upper-level population $N_2$ is dominated by the processes of spontaneous fluorescence emitting and possible energy transfer to silver particles, by contrary, the decay of lower-level population $N_1$ is mainly determined by intrinsic nonradiative losses. Thereby the associated rate equations of laser molecules are given by
\begin{subequations}
\begin{eqnarray}
\frac{{dN_1 }}{{dt}} &=&  - \gamma _1^{nr} \;(N_1  - N_1^{eq} ) + (N_2  - N_1 )B_{st} I(\omega ),\label{rateequation a}\\
\frac{{dN_2 }}{{dt}} &=& \lambda _p  - (\gamma _2^{abs} \; + A_{sp} )(N_2  - N_2^{eq} ) - (N_2  - N_1 )B_{st} I(\omega ),\label{rateequation b}
\end{eqnarray}
\end{subequations}
in which the pumping rate $\lambda_p$ for dye molecules is proportional to $\left| {{\bf{E}}_{\rm{p}}  \cdot {\bf{p}}_{{\rm{pump}}} } \right|^2$,
with $\bf{E}_{\rm{p}}$ being the local pump field and ${\bf{p}}_{{\rm{pump}}}$ the related transition dipole moment;
$I(\omega)$ is the partial energy intensity of local laser field that is 'seen' by the dye molecules;
$A_{sp}$ and $B_{st}$ are the Einstein coefficients corresponding to the spontaneous and stimulated emissions, respectively; the superscript 'eq' labels the thermal equilibrium values of the two populations; $\gamma _2^{abs}$ presents the absorption rate caused by Ag nanoparticles; and $\gamma _1^{nr}$ denotes the intrinsic nonradiative decay rate of lower-level dye molecules. Following rate equations (8), the steady population inversion is given by
\begin{equation}
\label{population inversion}
\Delta N = \frac{{\lambda _p (\gamma _2^{abs} \; + A_{sp} )^{ - 1}  - (N_1^{eq}  - N_2^{eq} )}}{{1 + \left[ {(\gamma _1^{nr} )^{ - 1}  + (\gamma _2^{abs} \; + A_{sp} )^{ - 1} } \right]B_{st} I(\omega )}},
\end{equation}
which in principle determines the threshold condition, $\Delta N = 0$, for the local net output of stimulated photons and predicts a gain saturation at high laser intensity. As making use of the distance-to-particle dependent pumping rate, energy absorption rate and spontaneous emission coefficient, proposed in a bunch of literatures\cite{chew1987transition,klimov2001spontaneous,anger2006enhancement,bharadwaj2007spectral},
for dipole fluorophores outside a metal nanoparticle, we can infer a positive population inversion for the dye molecules slightly separated from the Ag nanoparticle if under sufficient pumping. To clarify this property we need only to focus on the numerator of expression (9) and plot (in Figure 4) a radial profile of $\lambda _p (\gamma _2^{abs} \; + A_{sp} )^{ - 1}$ normalized by dividing its asymptotic value $\lambda _p^{(0)} (A_{sp}^{(0)} )^{ - 1}$ in far-field region, with $\lambda _p^{(0)}$ and $A_{sp}^{(0)}$ being denoted as the far-field limits of the pumping rate and spontaneous emission rate, respectively. It is worth noting that when the dye molecules close to the metallic surface, the enhancement of pumping rate is limited while the energy absorption rate becomes divergent, and thus \ref{population inversion} presents a minus population inversion, resulting in a laser quenching.
Because the net emission power of laser photons per unit volume is characterized by $\Delta N\,B_{st} \,I(\omega ) \propto \Delta N\left| {{\bf{E}} \cdot {\bf{p}}_{{\rm{emit}}} } \right|^2$ with $\bf{p}_{\rm{emit}}$ representing the stimulated transition dipole moment at laser frequencies, substituting \ref{near field} and \ref{population inversion} into this expression, we assert that at a vicinity of Ag nanoparticles the enlarged population inversion of laser dye, multiplied by the enhanced near-field triggering stimulated transitions, would create a substantial lasing. This mechanism is the key for understanding why coherent lasing resonances can so easily be archived in Ag-nanoparticle-based amplifying media.
\begin{figure}[h]
\begin{center}
\includegraphics[width=8.0cm]{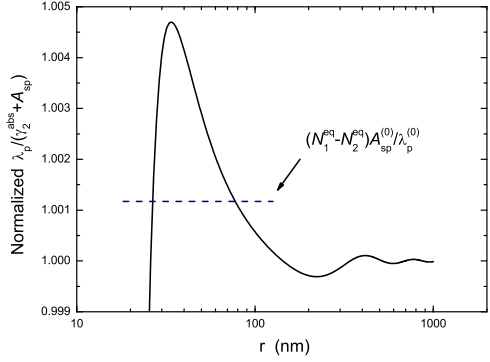}
\end{center}

\caption{Normalized $\lambda _p (\gamma _2^{abs} + A_{sp} )^{ - 1}$ curve, for dye molecules perpendicular to particle surface and parallel to the polarization direction of incident pumping light, indicates a population ripple around the particle.
The calculation of the total decay rate $(\gamma _2^{abs} + A_{sp})$  follows references \cite{chew1987transition,klimov2001spontaneous}. The dashed horizontal line refers to a pump-strength-dependent critical value $(N_1^{eq} - N_2^{eq} )A_{sp}^{(0)} (\lambda _p^{(0)} )^{ - 1}$, above which the curve corresponds to a positive population inversion.}
\label{population enhancement}
\end{figure}

Upon the pump energy is moderate and the intensity of the applied electric field surpasses 0.25 $V/ \mu m$, the lasing spectra exhibit an intriguing frequency-splitting, i.e., two remarkable spike-groups with a wavelength-separation of several nanometers appearing at the top of each emission spectrum. As shown in Figure 5a, a pair of maximum emission spikes that respectively belong to the two separated spike-groups are denoted by $P_1$ and $P_2$, in which the former almost always localizes at 610 nm, while the latter tends to settle at the vicinity of 607 nm and experiences a slightly red shift with increasing the external voltage. The insert in Figure 5a illustrates the evolution of the wavelength difference between $P_1$ and $P_2$ as a function of the applied voltage across the NLC-Ag-DCM sample. This frequency-splitting is likely due to the dielectric anisotropy of the NLC host, because the subtle difference between the polarizabilities parallel and vertical to the optical axis of a NLC host gives rise to different interface polarizations, which was found to discriminately screen the initial local fields of pure metal particle plasmons and thus split the plasmonic resonance frequencies\cite{park2004splitting,park2005surface}.

\begin{figure}[h]
\begin{center}
\includegraphics[width=8.0cm]{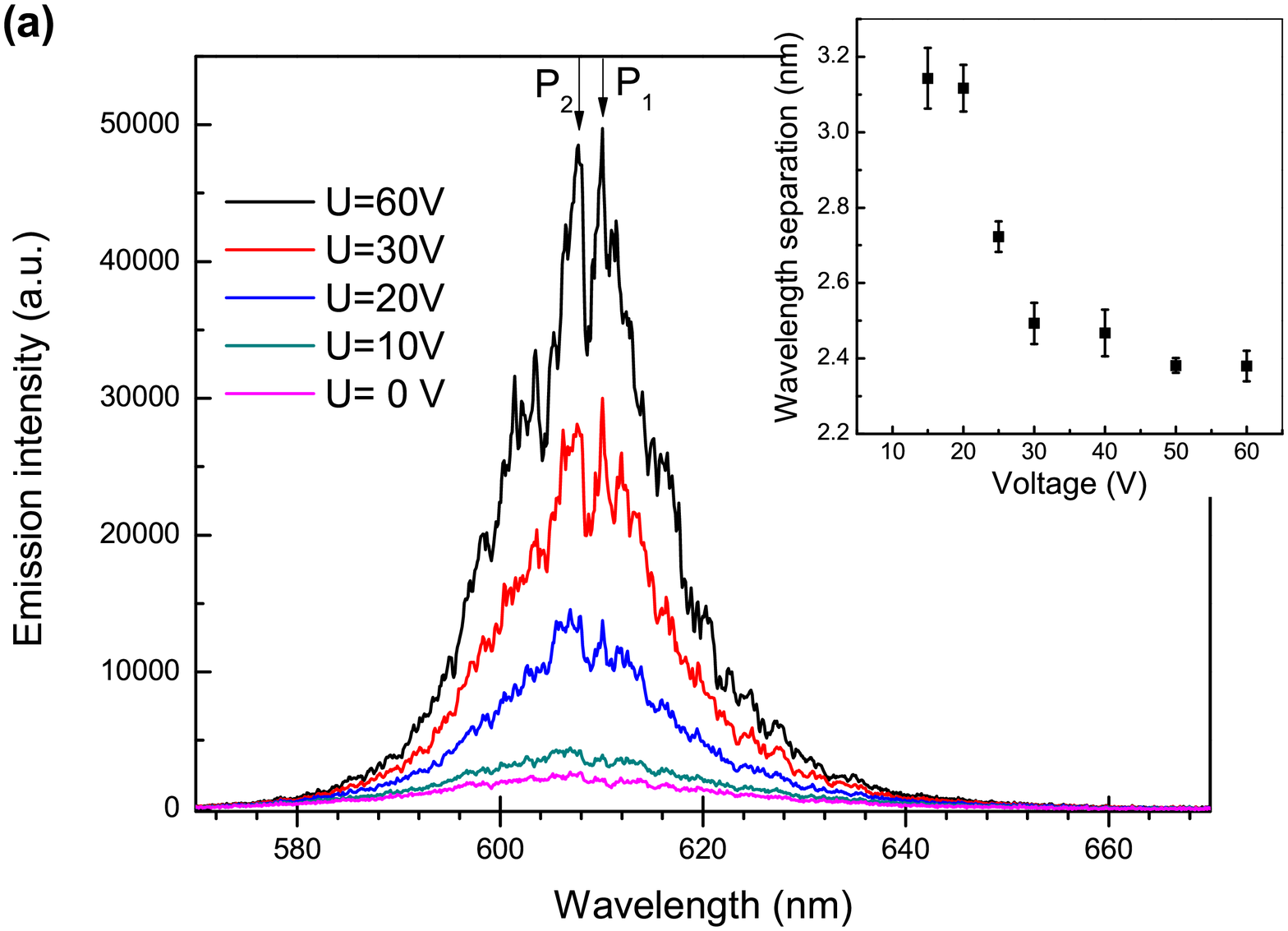}
\includegraphics[width=8.0cm]{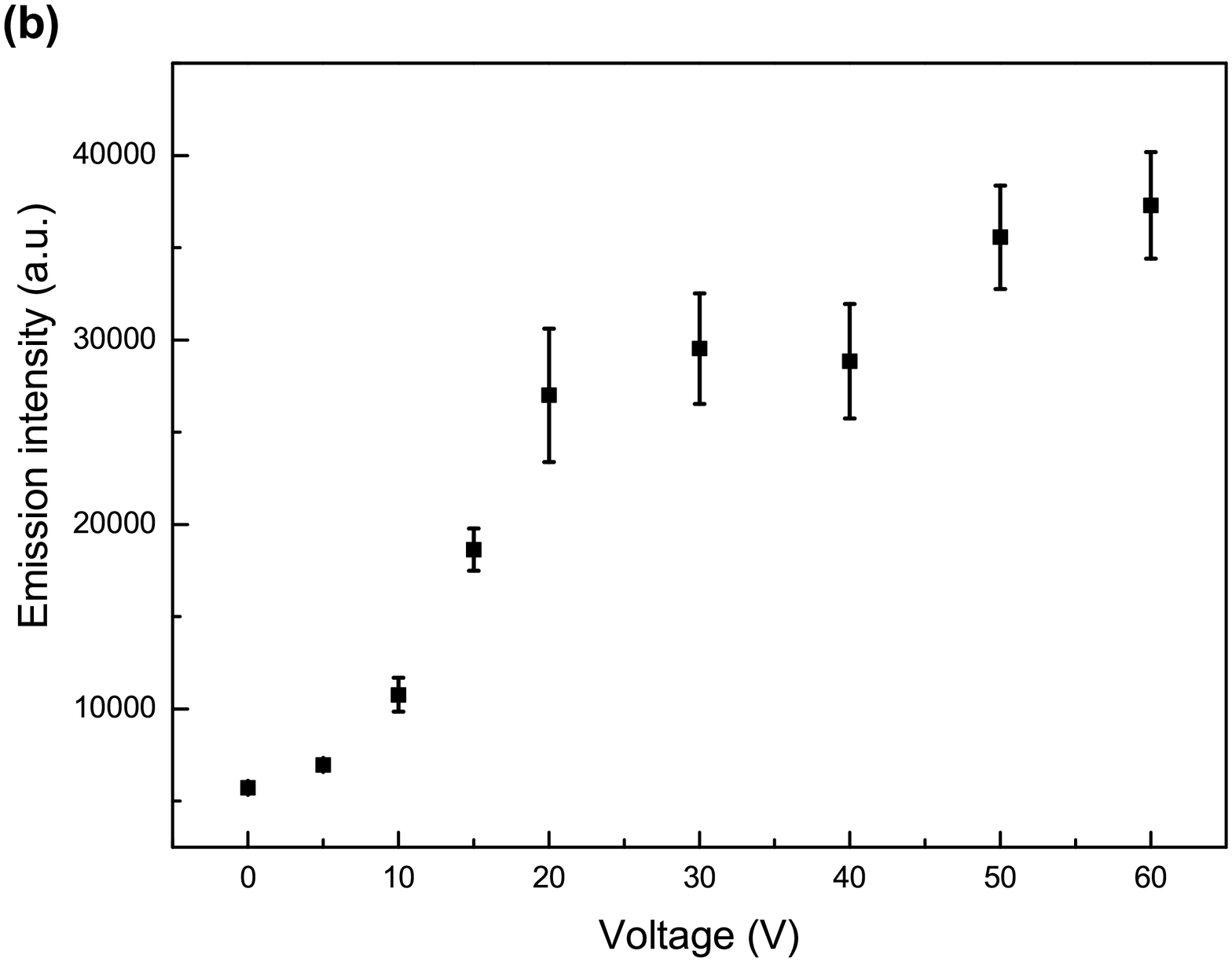}

\end{center}

\caption{(a) Main panel shows the electric field dependence of emission spectra, featured by striking frequency splitting under stronger applied fields, and the insert gives the corresponding wavelength separation as a function of applied transverse voltage. (b) Evolution of the emission intensity against the transverse voltage exhibits a switch behavior around 15 V. All experimental data were collected from a pumped strip with $\sim$1.8mm length and $\sim$0.20mm width, and error bars correspond to the standard deviations.}
\label{field dependence}
\end{figure}

To investigate the influence of the director fluctuation of the NLC host on the emission properties, we let the transverse electric field increase from 0 to 60 volts at a fixed pump strength. The evolution of the emission intensity against the transverse voltage is illustrated in Figure 5b.
At a weak electric field below 0.25 $V/\mu m$, we only obtained flat spontaneous emission spectra no matter how strong was the pumping. While as the applied electric field grew up to 0.5 $V/\mu m$, the emission spectra exhibited a pronounced random lasing property, and the peak of the emission intensity experienced a rapid climb. By continuously modulating the voltage upward, we found that the further increasing of the emission intensity become less significant and eventually negligible. This indicates that introducing dielectric perturbation into the host medium can effectively control the lasing resonances of the NLC-Ag-DCM sample.

In contrast to the concept that a robust and steady scattering mechanism is favorable to support coherent feedback for lasing resonances\cite{frolov1999stimulated,cao2003lasing,wiersma2008physics}, yet here the transient and dynamic one in the NLC medium actually disturbs coherent feedback and quenches lasing resonances. There is a simple understanding of the dependence of the lasing intensity on the external electric field, by looking into the scattering details of the NLC. The differential cross section of light scattering per unit
volume of the NLC is given by \cite{chandrasekhar1992}
\begin{equation}\label{scattering section}
\frac{{d\sigma }}{{d\Omega }} = \frac{{k^4 \varepsilon _\Delta ^2 }}{{16\pi ^2 }}\sum\limits_{\alpha  = x,y} {\left[ {({\bf{p}}_s  \cdot {\bf{e}}_\alpha  )({\bf{p}}_i  \cdot {\bf{e}}_z ) + ({\bf{p}}_i  \cdot {\bf{e}}_\alpha  )({\bf{p}}_s  \cdot {\bf{e}}_z )} \right]^2 I_\alpha  ({\bf{q}})} ,
\end{equation}
where ${\bf{q}} = {\bf{k}}_i  - {\bf{k}}_s $, with $ {\bf{k}}_i $ and $ {\bf{k}}_s $ representing the propagation vectors for any incident and scattered light, respectively; the unit vectors $ {\bf{p}}_i $ and $ {\bf{p}}_s $ characterize the polarization directions of the incident and scattered light, respectively; the longitudinal unit vector $ {\bf{e}}_z $, as matching the oriented electric field, dictates the average orientation of the NLC's director field $ {\bf{\tilde n}} \equiv {\bf{\tilde n}}(x,y,z,t) $ in thermal equilibrium; and $I_{x,y} ({\bf{q}})$ denote the time averages for the Fourier components of the spatial correlations of two transverse director fluctuations, which can be calculated from the following variation of the total Frank free energy \cite{gennes1993physics,gennes1968fluctuations,gennes1969Long}
\begin{equation}\label{Frank free energy}
\delta G = \int_V {\delta g_{bulk} d\tau }
\end{equation}
over scattering region. The Frank free energy density involved in such variation is given by
\begin{equation}\label{}
\begin{array}{l}
 g_{bulk}  = \frac{1}{2}\left[ {k_{11} \left( {\nabla  \cdot {\bf{\tilde n}}} \right)^2  + k_{22} \left( {{\bf{\tilde n}} \cdot \nabla  \times {\bf{\tilde n}}} \right)^2 } \right. \\
 \quad \quad \quad + k_{33} \left( {{\bf{\tilde n}} \times \nabla  \times {\bf{\tilde n}}} \right)^2  - \left. {\varepsilon _\Delta  \left( {{\bf{\tilde n}} \cdot {\bf{E}}_{ap} } \right)^2 } \right] + Const, \\
 \end{array}
\end{equation}
in which we add the last director-dependent term coming from the dielectric polarization caused by the applied electrical field ${\bf{E}}_{ap}$ along z-axis; $k_{11}$, $k_{22}$ and $k_{33}$ are the standard Frank elastic constants corresponding to splay, twist and bend deformations, respectively. The director field is assumed to deviate from the average direction by small fluctuations but conserve the length, i.e., ${\bf{\tilde n}} = {\bf{e}}_z  + \delta {\bf{\tilde n}} = (\delta \tilde n_x ,\delta \tilde n_y ,1 + \delta \tilde n_z ) $ and ${\bf{\tilde n}} \cdot {\bf{\tilde n}} = 1 $.
Starting from the \ref{Frank free energy}, following a similar procedure introduced in reference \cite{chandrasekhar1992}, and then utilizing the  energy equipartition theorem, we obtain
\begin{equation}\label{spatial correlation}
\begin{array}{l}
 I_x ({\bf{q}}) = k_B T\left( {k_{11} q_x^2  + k_{22} q_y^2  + k_{33} q_z^2  + \varepsilon _\Delta  E_{ap}^2 } \right)^{ - 1} , \\
 I_y ({\bf{q}}) = k_B T\left( {k_{11} q_y^2  + k_{22} q_x^2  + k_{33} q_z^2  + \varepsilon _\Delta  E_{ap}^2 } \right)^{ - 1} , \\
 \end{array}
\end{equation}
in which $k_B$ is the Boltzmann constant, and T the absolute temperature. Combining \ref{scattering section} and \ref{spatial correlation}, we see that the director fluctuations and, consequently, the scattering cross section $ {d\sigma }/{d\Omega } $ monotonically decline with increasing the applied electric field strength. It means that the dynamic dielectric perturbation in a NLC host can be really controlled by modulating external fields.

The fluctuated dielectric perturbations impact the optical properties of the NLC-Ag-DCM sample at least in two aspects: one is damaging the steady interferences among different particle plasmon-polariton fields, which washes away the coupled plasmonic resonances; the other is diminishing the total emission intensity of both coherent and incoherent lights by scattering events. Thus we relate the observed laser quenching to the damage of the coupled plasmonic resonances, and the dimming of total emission the dynamic light scattering, which removes part of photons out of the emission beam. Furthermore, the perish of lasing spikes at a low strength of the applied field suggests us that the multiple scattering mechanism, originating from the simple thermal fluctuation of NLC director orientation, fails to create sufficient coherent feedback to develop lasing resonances, under the present pumping condition of nanosecond pulses if not in more general circumstances.

\subsection{Conclusions}
In summary, we demonstrated an electrical control over plasmonic random lasing resonances which had never been performed previously. The metal nanoparticles distributed in the gain medium of dye-doped NLC contribute to a strong local-field enhancement on the stimulated emission, while the dynamic light scattering in the NLC material, essentially a result of the thermal fluctuation of the director orientation and the dynamic dissipation due to NLC's viscosity, leads to broadening and quenching the lasing resonances. The switch behavior on the random lasing roots in the competition between these two opposite mechanisms, which can be changed by modulating the director fluctuation strength with applied electric fields. The disappearance of lasing resonances in the absence of applied fields also means that the intrinsic multiple dynamic light scattering of NLC materials hardly provides coherent optical feedback for random lasing. As an ultimate goal to accurately predict and design the lasing resonances for general amplifying media modified by metal nanoparticles, further investigations are on the horizon.

\bibliography{achemso-demo}

\providecommand*{\mcitethebibliography}{\thebibliography}
\csname @ifundefined\endcsname{endmcitethebibliography}
{\let\endmcitethebibliography\endthebibliography}{}
\begin{mcitethebibliography}{33}
\providecommand*{\natexlab}[1]{#1}
\providecommand*{\mciteSetBstSublistMode}[1]{}
\providecommand*{\mciteSetBstMaxWidthForm}[2]{}
\providecommand*{\mciteBstWouldAddEndPuncttrue}
  {\def\EndOfBibitem{\unskip.}}
\providecommand*{\mciteBstWouldAddEndPunctfalse}
  {\let\EndOfBibitem\relax}
\providecommand*{\mciteSetBstMidEndSepPunct}[3]{}
\providecommand*{\mciteSetBstSublistLabelBeginEnd}[3]{}
\providecommand*{\EndOfBibitem}{}
\mciteSetBstSublistMode{f}
\mciteSetBstMaxWidthForm{subitem}{(\alph{mcitesubitemcount})}
\mciteSetBstSublistLabelBeginEnd{\mcitemaxwidthsubitemform\space}
{\relax}{\relax}

\bibitem[Lawandy et~al.(1994)Lawandy, Balachandran, Gomes, and
  Sauvain]{lawandy1994laser}
Lawandy,~N.~M.; Balachandran,~R.; Gomes,~A.; Sauvain,~E. \emph{Nature}
  \textbf{1994}, \emph{368}, 436--438\relax
\mciteBstWouldAddEndPuncttrue
\mciteSetBstMidEndSepPunct{\mcitedefaultmidpunct}
{\mcitedefaultendpunct}{\mcitedefaultseppunct}\relax
\EndOfBibitem
\bibitem[Cao et~al.(1999)Cao, Zhao, Ho, Seelig, Wang, and Chang]{cao1999random}
Cao,~H.; Zhao,~Y.; Ho,~S.; Seelig,~E.; Wang,~Q.; Chang,~R. \emph{Physical
  Review Letters} \textbf{1999}, \emph{82}, 2278--2281\relax
\mciteBstWouldAddEndPuncttrue
\mciteSetBstMidEndSepPunct{\mcitedefaultmidpunct}
{\mcitedefaultendpunct}{\mcitedefaultseppunct}\relax
\EndOfBibitem
\bibitem[Garc{\'\i}a et~al.(2007)Garc{\'\i}a, Sapienza, Blanco, and
  L{\'o}pez]{garcia2007photonic}
Garc{\'\i}a,~P.~D.; Sapienza,~R.; Blanco,~{\'A}.; L{\'o}pez,~C. \emph{Advanced
  Materials} \textbf{2007}, \emph{19}, 2597--2602\relax
\mciteBstWouldAddEndPuncttrue
\mciteSetBstMidEndSepPunct{\mcitedefaultmidpunct}
{\mcitedefaultendpunct}{\mcitedefaultseppunct}\relax
\EndOfBibitem
\bibitem[Noginov(2005)]{Noginov2005lasers}
Noginov,~M.~A. \emph{Solid-State Random Lasers};
\newblock Springer: Berlin, 2005;
\newblock Vol. 105\relax
\mciteBstWouldAddEndPuncttrue
\mciteSetBstMidEndSepPunct{\mcitedefaultmidpunct}
{\mcitedefaultendpunct}{\mcitedefaultseppunct}\relax
\EndOfBibitem
\bibitem[Meng et~al.(2008)Meng, Fujita, Zong, Murai, and
  Tanaka]{meng2008random}
Meng,~X.; Fujita,~K.; Zong,~Y.; Murai,~S.; Tanaka,~K. \emph{Applied Physics
  Letters} \textbf{2008}, \emph{92}, 201112\relax
\mciteBstWouldAddEndPuncttrue
\mciteSetBstMidEndSepPunct{\mcitedefaultmidpunct}
{\mcitedefaultendpunct}{\mcitedefaultseppunct}\relax
\EndOfBibitem
\bibitem[Meng et~al.(2009)Meng, Fujita, Murai, Zong, Akasaka, Hasegawa, and
  Tanaka]{meng2009coherent}
Meng,~X.; Fujita,~K.; Murai,~S.; Zong,~Y.; Akasaka,~S.; Hasegawa,~H.;
  Tanaka,~K. \emph{physica status solidi (c)} \textbf{2009}, \emph{6},
  S102--S105\relax
\mciteBstWouldAddEndPuncttrue
\mciteSetBstMidEndSepPunct{\mcitedefaultmidpunct}
{\mcitedefaultendpunct}{\mcitedefaultseppunct}\relax
\EndOfBibitem
\bibitem[de~Gennes and Prost(1993)]{gennes1993physics}
de~Gennes,~P.~G.; Prost,~J. \emph{The Physics of Liquid Crystals};
\newblock International Series of Monographs on Physics;
\newblock Clarendon Press: Oxford, 1993\relax
\mciteBstWouldAddEndPuncttrue
\mciteSetBstMidEndSepPunct{\mcitedefaultmidpunct}
{\mcitedefaultendpunct}{\mcitedefaultseppunct}\relax
\EndOfBibitem
\bibitem[Shankar et~al.(2010)Shankar, Wu, and Bigioni]{shankar2010wet}
Shankar,~R.; Wu,~B.~B.; Bigioni,~T.~P. \emph{The Journal of Physical Chemistry
  C} \textbf{2010}, \emph{114}, 15916--15923\relax
\mciteBstWouldAddEndPuncttrue
\mciteSetBstMidEndSepPunct{\mcitedefaultmidpunct}
{\mcitedefaultendpunct}{\mcitedefaultseppunct}\relax
\EndOfBibitem
\bibitem[John(1984)]{john1984electromagnetic}
John,~S. \emph{Phys. Rev. Lett.} \textbf{1984}, \emph{53}, 2169--2172\relax
\mciteBstWouldAddEndPuncttrue
\mciteSetBstMidEndSepPunct{\mcitedefaultmidpunct}
{\mcitedefaultendpunct}{\mcitedefaultseppunct}\relax
\EndOfBibitem
\bibitem[Anderson(1985)]{Anderson1985}
Anderson,~P. \emph{Phil. Mag. B} \textbf{1985}, \emph{52}, 505--508\relax
\mciteBstWouldAddEndPuncttrue
\mciteSetBstMidEndSepPunct{\mcitedefaultmidpunct}
{\mcitedefaultendpunct}{\mcitedefaultseppunct}\relax
\EndOfBibitem
\bibitem[Lamprecht et~al.(2000)Lamprecht, Schider, Lechner, Ditlbacher, Krenn,
  Leitner, Aussenegg,et~al.]{lamprecht2000metal}
Lamprecht,~B.; Schider,~G.; Lechner,~R.; Ditlbacher,~H.; Krenn,~J.;
  Leitner,~A.; Aussenegg,~F. et~al.Lamprecht, B and Schider, G and Lechner, RT
  and Ditlbacher, H and Krenn, JR and Leitner, A and Aussenegg, FR and others
  \emph{Physical review letters} \textbf{2000}, \emph{84}, 4721--4724\relax
\mciteBstWouldAddEndPuncttrue
\mciteSetBstMidEndSepPunct{\mcitedefaultmidpunct}
{\mcitedefaultendpunct}{\mcitedefaultseppunct}\relax
\EndOfBibitem
\bibitem[Quinten(2001)]{quinten2001color}
Quinten,~M. \emph{Applied Physics B} \textbf{2001}, \emph{73}, 317--326\relax
\mciteBstWouldAddEndPuncttrue
\mciteSetBstMidEndSepPunct{\mcitedefaultmidpunct}
{\mcitedefaultendpunct}{\mcitedefaultseppunct}\relax
\EndOfBibitem
\bibitem[Su et~al.(2003)Su, Wei, Zhang, Mock, Smith, and
  Schultz]{su2003interparticle}
Su,~K.-H.; Wei,~Q.-H.; Zhang,~X.; Mock,~J.; Smith,~D.~R.; Schultz,~S.
  \emph{Nano Letters} \textbf{2003}, \emph{3}, 1087--1090\relax
\mciteBstWouldAddEndPuncttrue
\mciteSetBstMidEndSepPunct{\mcitedefaultmidpunct}
{\mcitedefaultendpunct}{\mcitedefaultseppunct}\relax
\EndOfBibitem
\bibitem[Nordlander et~al.(2004)Nordlander, Oubre, Prodan, Li, and
  Stockman]{nordlander2004plasmon}
Nordlander,~P.; Oubre,~C.; Prodan,~E.; Li,~K.; Stockman,~M. \emph{Nano Letters}
  \textbf{2004}, \emph{4}, 899--903\relax
\mciteBstWouldAddEndPuncttrue
\mciteSetBstMidEndSepPunct{\mcitedefaultmidpunct}
{\mcitedefaultendpunct}{\mcitedefaultseppunct}\relax
\EndOfBibitem
\bibitem[Fromm et~al.(2004)Fromm, Sundaramurthy, Schuck, Kino, and
  Moerner]{fromm2004gap}
Fromm,~D.~P.; Sundaramurthy,~A.; Schuck,~P.~J.; Kino,~G.; Moerner,~W.
  \emph{Nano Letters} \textbf{2004}, \emph{4}, 957--961\relax
\mciteBstWouldAddEndPuncttrue
\mciteSetBstMidEndSepPunct{\mcitedefaultmidpunct}
{\mcitedefaultendpunct}{\mcitedefaultseppunct}\relax
\EndOfBibitem
\bibitem[Romero et~al.(2006)Romero, Aizpurua, Bryant, and Garc{\'\i}a
  De~Abajo]{romero2006plasmons}
Romero,~I.; Aizpurua,~J.; Bryant,~G.~W.; Garc{\'\i}a De~Abajo,~F.~J.
  \emph{Optics Express} \textbf{2006}, \emph{14}, 9988--9999\relax
\mciteBstWouldAddEndPuncttrue
\mciteSetBstMidEndSepPunct{\mcitedefaultmidpunct}
{\mcitedefaultendpunct}{\mcitedefaultseppunct}\relax
\EndOfBibitem
\bibitem[Biring et~al.(2008)Biring, Wang, Wang, and Wang]{biring2008light}
Biring,~S.; Wang,~H.-H.; Wang,~J.-K.; Wang,~Y.-L. \emph{Optics express}
  \textbf{2008}, \emph{16}, 15312--15324\relax
\mciteBstWouldAddEndPuncttrue
\mciteSetBstMidEndSepPunct{\mcitedefaultmidpunct}
{\mcitedefaultendpunct}{\mcitedefaultseppunct}\relax
\EndOfBibitem
\bibitem[Bohren and Huffman(1998)]{bohren1998absorption}
Bohren,~C.~F.; Huffman,~D.~R. \emph{Absorption and scattering of light by small
  particles};
\newblock John Wiley $\&$ Sons, 1998\relax
\mciteBstWouldAddEndPuncttrue
\mciteSetBstMidEndSepPunct{\mcitedefaultmidpunct}
{\mcitedefaultendpunct}{\mcitedefaultseppunct}\relax
\EndOfBibitem
\bibitem[Johnson and Christy(1972)]{johnson1972optical}
Johnson,~P.; Christy,~R. \emph{Physical Review B} \textbf{1972}, \emph{6},
  4370--4379\relax
\mciteBstWouldAddEndPuncttrue
\mciteSetBstMidEndSepPunct{\mcitedefaultmidpunct}
{\mcitedefaultendpunct}{\mcitedefaultseppunct}\relax
\EndOfBibitem
\bibitem[Boyd(2003)]{boyd2003nonlinear}
Boyd,~R.~W. \emph{Nonlinear optics}, Second edition ed.;
\newblock Academic press: London, 2003\relax
\mciteBstWouldAddEndPuncttrue
\mciteSetBstMidEndSepPunct{\mcitedefaultmidpunct}
{\mcitedefaultendpunct}{\mcitedefaultseppunct}\relax
\EndOfBibitem
\bibitem[Chew(1987)]{chew1987transition}
Chew,~H. \emph{The Journal of chemical physics} \textbf{1987}, \emph{87},
  1355--1360\relax
\mciteBstWouldAddEndPuncttrue
\mciteSetBstMidEndSepPunct{\mcitedefaultmidpunct}
{\mcitedefaultendpunct}{\mcitedefaultseppunct}\relax
\EndOfBibitem
\bibitem[Klimov et~al.(2001)Klimov, Ducloy, and
  Letokhov]{klimov2001spontaneous}
Klimov,~V.~V.; Ducloy,~M.; Letokhov,~V.~S. \emph{Quantum Electronics}
  \textbf{2001}, \emph{31}, 569--586\relax
\mciteBstWouldAddEndPuncttrue
\mciteSetBstMidEndSepPunct{\mcitedefaultmidpunct}
{\mcitedefaultendpunct}{\mcitedefaultseppunct}\relax
\EndOfBibitem
\bibitem[Anger and Bharadwaj(2006)]{anger2006enhancement}
Anger,~P.; Bharadwaj,~P. \emph{Physical review letters} \textbf{2006},
  \emph{96}, 113002\relax
\mciteBstWouldAddEndPuncttrue
\mciteSetBstMidEndSepPunct{\mcitedefaultmidpunct}
{\mcitedefaultendpunct}{\mcitedefaultseppunct}\relax
\EndOfBibitem
\bibitem[Bharadwaj and Novotny(2007)]{bharadwaj2007spectral}
Bharadwaj,~P.; Novotny,~L. \emph{Optics Express} \textbf{2007}, \emph{15},
  14266--14274\relax
\mciteBstWouldAddEndPuncttrue
\mciteSetBstMidEndSepPunct{\mcitedefaultmidpunct}
{\mcitedefaultendpunct}{\mcitedefaultseppunct}\relax
\EndOfBibitem
\bibitem[Park and Stroud(2004)]{park2004splitting}
Park,~S.~Y.; Stroud,~D. \emph{Applied physics letters} \textbf{2004},
  \emph{85}, 2920--2922\relax
\mciteBstWouldAddEndPuncttrue
\mciteSetBstMidEndSepPunct{\mcitedefaultmidpunct}
{\mcitedefaultendpunct}{\mcitedefaultseppunct}\relax
\EndOfBibitem
\bibitem[Park and Stroud(2005)]{park2005surface}
Park,~S.~Y.; Stroud,~D. \emph{Physical review letters} \textbf{2005},
  \emph{94}, 217401\relax
\mciteBstWouldAddEndPuncttrue
\mciteSetBstMidEndSepPunct{\mcitedefaultmidpunct}
{\mcitedefaultendpunct}{\mcitedefaultseppunct}\relax
\EndOfBibitem
\bibitem[Frolov et~al.(1999)Frolov, Vardeny, Yoshino, Zakhidov, and
  Baughman]{frolov1999stimulated}
Frolov,~S.; Vardeny,~Z.; Yoshino,~K.; Zakhidov,~A.; Baughman,~R. \emph{Physical
  Review B} \textbf{1999}, \emph{59}, R5284\relax
\mciteBstWouldAddEndPuncttrue
\mciteSetBstMidEndSepPunct{\mcitedefaultmidpunct}
{\mcitedefaultendpunct}{\mcitedefaultseppunct}\relax
\EndOfBibitem
\bibitem[Cao(2003)]{cao2003lasing}
Cao,~H. \emph{Waves in random media} \textbf{2003}, \emph{13}, R1--R39\relax
\mciteBstWouldAddEndPuncttrue
\mciteSetBstMidEndSepPunct{\mcitedefaultmidpunct}
{\mcitedefaultendpunct}{\mcitedefaultseppunct}\relax
\EndOfBibitem
\bibitem[Wiersma(2008)]{wiersma2008physics}
Wiersma,~D.~S. \emph{Nature Physics} \textbf{2008}, \emph{4}, 359--367\relax
\mciteBstWouldAddEndPuncttrue
\mciteSetBstMidEndSepPunct{\mcitedefaultmidpunct}
{\mcitedefaultendpunct}{\mcitedefaultseppunct}\relax
\EndOfBibitem
\bibitem[Chandrasekhar(1992)]{chandrasekhar1992}
Chandrasekhar,~S. \emph{Liquid Crystals}, 2nd ed.;
\newblock Cambridge University Press, 1992\relax
\mciteBstWouldAddEndPuncttrue
\mciteSetBstMidEndSepPunct{\mcitedefaultmidpunct}
{\mcitedefaultendpunct}{\mcitedefaultseppunct}\relax
\EndOfBibitem
\bibitem[de~Gennes(1968)]{gennes1968fluctuations}
de~Gennes,~P.~G. \emph{C. R. Acad. Sci. Paris.} \textbf{1968}, \emph{266},
  15--17\relax
\mciteBstWouldAddEndPuncttrue
\mciteSetBstMidEndSepPunct{\mcitedefaultmidpunct}
{\mcitedefaultendpunct}{\mcitedefaultseppunct}\relax
\EndOfBibitem
\bibitem[de~Gennes(1969)]{gennes1969Long}
de~Gennes,~P.~G. \emph{Mol. Cryst. $\&$ Liq. Cryst.} \textbf{1969}, \emph{7},
  325--345\relax
\mciteBstWouldAddEndPuncttrue
\mciteSetBstMidEndSepPunct{\mcitedefaultmidpunct}
{\mcitedefaultendpunct}{\mcitedefaultseppunct}\relax
\EndOfBibitem
\end{mcitethebibliography}

\end{spacing}

\end{document}